\documentclass[12pt,fleqn,aps,prd,showpacs,nofootinbib]{revtex4}
\usepackage{epsf}
\usepackage{amsfonts,amssymb,hhline}
\def\cm{\hspace*{1cm}}
\def\inch{\hspace*{1in}}

\def\noi{\noindent}

\def\Jl#1#2{#1 {\bf #2},\ }

\def\ApJ#1 {\Jl{Astroph. J.}{#1}}
\def\CQG#1 {\Jl{Class. Quantum Grav.}{#1}}
\def\DAN#1 {\Jl{Dokl. AN SSSR}{#1}}
\def\GC#1 {\Jl{Grav. Cosmol.}{#1}}
\def\GRG#1 {\Jl{Gen. Rel. Grav.}{#1}}
\def\JETF#1 {\Jl{Zh. Eksp. Teor. Fiz.}{#1}}
\def\JETP#1 {\Jl{Sov. Phys. JETP}{#1}}
\def\JHEP#1 {\Jl{JHEP}{#1}}
\def\JMP#1 {\Jl{J. Math. Phys.}{#1}}
\def\NPB#1 {\Jl{Nucl. Phys.}{B\ #1}}
\def\NP#1 {\Jl{Nucl. Phys.}{#1}}
\def\PLA#1 {\Jl{Phys. Lett.}{#1A}}
\def\PLB#1 {\Jl{Phys. Lett.}{#1B}}
\def\PRD#1 {\Jl{Phys. Rev.}{D\ #1}}
\def\PRL#1 {\Jl{Phys. Rev. Lett.}{#1}}

\def\al{&}
\def\lal{&& {}}
\def\eq{Eq.\,}
\def\eqs{Eqs.\,}
\def\beq{\begin{equation}}
\def\eeq{\end{equation}}
\def\bear{\begin{eqnarray}}
\def\bearr{\begin{eqnarray} \lal}
\def\ear{\end{eqnarray}}
\def\earn{\nonumber \end{eqnarray}}
\def\nn{\nonumber\\ {}}

\def\nnn{\nonumber\\ \lal }

\def\eql{\al =\al}

\def\dst{\displaystyle}
\def\tst{\textstyle}
\def\fracd#1#2{{\dst\frac{#1}{#2}}}
\def\fract#1#2{{\tst\frac{#1}{#2}}}
\def\Half{{\fracd{1}{2}}}

\def\e{{\,\rm e}}
\def\d{\partial}

\def\diag{\mathop{\rm diag}\nolimits}

\def\const{{\rm const}}
\def\eps{\varepsilon}
\def\ep{\epsilon}

\newcommand{\vars}[1]{\left\{\begin{array}{ll}#1\end{array}\right.}

\def\mn{_{\mu\nu}}
\def\mN{_{\mu}^{\nu}}
\def\MN{^{\mu\nu}}

\def\vac{{}_{\rm (vac)}}

\def\N{{\mathbb N}}

\def\tZ{{\widetilde Z}}

\def\sph{spherically symmetric}
\def\ssph{static, spherically symmetric}
\def\bh{black hole}
\def\bhs{black holes}
\def\KS{Kan\-tow\-ski-Sachs}

\begin{document}

\title{Horizons vs. singularities in spherically symmetric space-times}

\author{K.A. Bronnikov}
\affiliation{Center for Gravitation and Fundamental Metrology, VNIIMS,
     46 Ozyornaya St., Moscow 119361, Russia;\\
 Institute of Gravitation and Cosmology, PFUR, 6 Miklukho-Maklaya St.,
     Moscow 117198, Russia} \email {kb20@yandex.ru}

\author{E. Elizalde}
\affiliation{Consejo Superior de Investigaciones Cient\'{\i}ficas
        ICE/CSIC-IEEC Campus UAB, Facultat de Ci\`encies, Torre C5-Parell-2a
    pl, E-08193 Bellaterra (Barcelona) Spain}
    \email {elizalde@ieec.uab.es, elizalde@math.mit.edu}

\author{S.D. Odintsov}
\affiliation{Instituci\`o Catalana de Recerca i Estudis Avan\c{c}ats (ICREA)
    and Institut de Ci\`encies de l'Espai (IEEC-CSIC),
    Campus UAB, Facultat de Ci\`encies, Torre C5-Parell-2a pl,
    E-08193 Bellaterra (Barcelona) Spain;\\
    Laboratory of Fundamental Studies, Tomsk State Pedagogical University,
    Tomsk} \email {odintsov@ieec.uab.es}

\author{O.B. Zaslavskii}
\affiliation{Astronomical Institute of Kharkov V.N. Karazin National
    University, 35 Sumskaya St., Kharkov, 61022, Ukraine}
    \email {ozaslav@kharkov.ua}

\begin{abstract}
  We discuss different kinds of Killing horizons possible in \ssph\
  configurations and recently classified as ``usual'', ``naked'' and ``truly
  naked'' ones depending on the near-horizon behavior of transverse tidal
  forces acting on an extended body. We obtain necessary conditions for the
  metric to be extensible beyond a horizon in terms of an arbitrary radial
  coordinate and show that all truly naked horizons, as well as many of
  those previously characterized as naked and even usual ones, do not admit
  an extension and therefore must be considered as singularities. Some
  examples are given, showing which kinds of matter are able to create
  specific space-times with different kinds of horizons, including truly
  naked ones. Among them are fluids with negative pressure and scalar
  fields with a particular behavior of the potential. We also discuss
  horizons and singularities in \KS\ \sph\ cosmologies and present horizon
  regularity conditions in terms of an arbitrary time coordinate and proper
  (synchronous) time. It turns out that horizons of orders 2 and higher
  occur in infinite proper times in the past or future, but one-way
  communication with regions beyond such horizons is still possible.
\end{abstract}

%%\keywords{Black holes, singularities, anisotropic cosmology}
\pacs{04.70Bw, 04.40 Nr, 98.80.-k}

\maketitle

\section{Introduction}

  As was pointed out by Horowitz and Ross \cite{HR1, HR2}, in some static
  \bh\ space-times, although all curvature invariants are everywhere finite,
  including the event horizon, an extended body falling into the \bh\
  experiences enormous tidal forces in the immediate vicinity of the
  horizon. Large tidal forces emerge because in some cases the curvature
  components in a freely falling reference frame are significantly enhanced
  as compared with their static values. These objects were termed [1, 2]
  ``naked \bhs'' because the behavior of tidal forces at the horizon
  resembles that near naked singularities. Examples of this behavior have
  been found in a wide class of theories of gravity including
  supergravities that arise in the low-energy limit of string theory.

  Refs. \cite{ZP, Z} further classified such \bhs\ and the corresponding
  event horizons as simply ``naked'' and ``truly naked'' ones: in the
  former, tidal forces experienced by a freely falling body are enhanced
  with respect to the static frame but remain finite. In the latter, the
  tidal forces become infinite. These properties depend on the asymptotic
  behavior of the curvature tensor near the horizon.

  It should be noted that, despite the same physical idea behind the notion
  of a naked horizon in Refs.\,\cite{HR1, HR2} and \cite{ZP, Z}, the
  particular criteria are different. Namely, Refs.\,\cite{HR1, HR2}
  considered a number of examples of \bh\ solutions of dilaton gravity,
  supergravities etc. and showed that tidal forces near the horizon can
  become arbitrarily large as one approaches some (singular) limit in the
  parameter space of the corresponding solution, whereas the curvature
  invariants remain finite and become infinite only when the above limit is
  reached. The criterion used in \cite{ZP, Z} is more formal: a horizon is
  called naked if some quantity, characterizing the tidal forces at the
  horizon, is zero in the static reference frame but is finite and nonzero
  in a freely falling reference frame. (Such horizons, from the viewpoint of
  \cite{HR1, HR2}, may be characterized as ``potentially naked'' since
  the tidal forces can really become very large at some values of the
  solution parameters.) This definition is much more convenient for general
  studies and classification of horizons in different \bh\ solutions, and,
  using it, horizons in \ssph\ \bh\ metrics were divided in \cite{ZP, Z}
  into three classes: usual, naked and truly naked. The classification was
  formulated in terms of the near-horizon behavior of the metric in the
  curvature (Schwarzschild-like) coordinates.

  In the present paper, we study the extensibility of the metrics beyond
  horizons of different kinds. We show that the metric can be analytically
  (or at least sufficiently smoothly) extended under some stringent
  conditions which are explicitly written. All \bh\ metrics, for which the
  extensions are well-known, certainly satisfy these conditions. It turns
  out, however, that all ``truly naked'' horizons, as well as many of those
  previously characterized as naked and even usual ones, do not admit an
  extension and therefore must be considered as singularities. Indeed,
  timelike and null geodesic incompleteness is generally regarded as a
  criterion for the presence of a singularity (see, e.g., \cite{wald}),
  used, in particular, in the well-known singularity theorems. And it is
  precisely such incompleteness that takes place at inextensible horizons
  (sometimes also called ``singular horizons''), despite finite values of
  the curvature invariants.\footnote
    {For convenience, we still use the words ``horizon'' or ``truly
    naked horizon'' (TNH) in all such cases without a risk of confusion.
    }
  Our approach is thus somewhat complementary to the discussion on
  singularities in black hole physics conducted in Refs. \cite{tipler,pois,
  brady,nolan,ori}

  It is natural to extend the consideration to the cosmological counterparts
  of \ssph\ configurations, i.e., \KS (KS) cosmological models. It can be
  mentioned that KS cosmologies are not excluded by modern observations if
  one assumes their sufficiently early isotropization \cite{craw}, and
  the latter may follow from the process of matter creation from vacuum ---
  see a discussion and some estimates in \cite{bd07, bz1}. It is also known
  \cite{bz1} that if the matter content of the Universe satisfies the Null
  Energy Condition, then the only way of avoiding a cosmological singularity
  in the past of a KS cosmology is the beginning of the cosmological
  evolution from a Killing horizon. Thus there is a good reason for studying
  the properties of such horizons, and some of them are described here.

  The structure of the paper is as follows. Sec. II contains some general
  relations for \sph\ metrics, needed in the further study. Sec. III
  discusses the properties of the so-called quasiglobal coordinate as a
  convenient tool for studying the extensibility problem. In Sec. IV we
  obtain the horizon extensibility conditions in terms of an arbitrary
  radial coordinate and, using the curvature coordinate $r$, compare the
  extensibility criterion with the previously given \cite{ZP, Z}
  classification of horizons as usual, naked and truly naked ones
  in \ssph\ space-times. Sec. V contains a number of examples showing which
  kinds of matter can create configurations with different kinds of
  horizons, including TNHs. In Sec. VI we discuss horizons in KS cosmologies
  and present the extensibility conditions in terms of an arbitrary time
  coordinate and proper (synchronous) time. Sec. VII contains some
  concluding remarks, in particular, on different kinds of singularities
  in \sph\ configurations of matter with negative pressure.

\section{Geometry}

  The general \ssph\ metric can be written in the form\footnote
    {We use the metric signature $(+\ -\ -\ -)$ and the
    units in which $c = \hbar = G = 1$.}
\beq                                                       \label{ds}  % 1
     ds^2 = \e^{2\gamma(R)} dt^2 - \e^{2\alpha(R)} dR^2 - r^2(R) d\Omega^2,
    \cm
         d\Omega^2 \equiv d \theta^2 + \sin^2 \theta d\phi^2,
\eeq
  where $R$ is an arbitrary radial coordinate. The most frequently used
  curvature (Schwarzschild) coordinate corresponds to the ``gauge''
  condition $R \equiv r$. Another useful choice is the so-called quasiglobal
  radial coordinate $R=u$, corresponding to the ``gauge'' $g_{00}g_{11} =
  -1$, so that the metric is
\beq                                                        \label{ds-u}%2
      ds^2 = A(u) dt^2 - \frac{du^2}{A(u)} - r^2(u) d\Omega^2.
\eeq

  The nonzero components of the Riemann tensor in the above two gauges are
\bear                                                       \label{Riem}%3
     R^{01}{}_{01} \eql -\Half A'' = -\e^{-\alpha-\gamma}
            \left(\gamma_r \e^{\gamma-\alpha}\right)_r = -K_1;
\nn
     R^{02}{}_{02} \eql -\Half \frac{A'r'}{r}
            = -\e^{-2\alpha}\frac{\gamma_r}{r} = -K_2 = R^{03}{}_{03};
\nn
     R^{12}{}_{12} \eql - A\frac{r''}{r}- \Half \frac{A'r'}{r}
        = \e^{-2\alpha} \frac{\alpha_r}{r} = -K_3 = R^{13}{}_{13};
\nn
     R^{23}{}_{23} \eql \frac{1}{r^2} (1- Ar'{}^2)
            = \frac{1}{r^2}(\e^{-2\alpha}-1) = - K_4,
\ear
  where expressions for the metric (\ref{ds-u}) are given in each line
  after the first equality sign and those for the curvature coordinates
  [\eq (\ref{ds}) with $R\equiv r$] after the second equality sign. The prime
  stands for $d/du$ and the subscript $r$ for $d/dr$.

  As follows from the geodesic deviation equations, the tidal forces
  experienced by bodies in the gravitational field (\ref{ds}) are
  conveniently characterized by the quantities \cite{Z}
\beq
      Z := R^{12}{}_{12} - R^{02}{}_{02} = K_2 - K_3           \label{Z}
\eeq
  in the static reference frame and
\beq
      \tZ \sim \e^{-2\gamma} Z = Z/A                          \label{tZ}
\eeq
  in a freely falling reference frame near a horizon. These quantities
  have been used in \cite{Z} to distinguish usual, naked and truly naked
  horizons.

  Assuming $A < 0$ in the metric (\ref{ds-u}), we obtain a \KS\ (KS) \sph\
  cosmology in which $u$ is a quasiglobal time coordinate and $t$ acquires a
  spatial nature.  Let us re-denote $A \to -A$ and $t \to x$, so that
\beq
    ds^2 = \frac{du^2}{A(u)} - A(u) dx^2 - r^2(u) d\Omega^2.  \label{KS-u}
\eeq

  A natural time variable in cosmology is the proper (or synchronous) time
  $\tau$. If we use it, the KS metric reads
\beq
     ds^2 = d\tau^2 - a^2(\tau) dx^2 - r^2(\tau) d\Omega^2.   \label{KS-tau}
\eeq

  In terms of the quasiglobal coordinate $u$, expressions for the nonzero
  Riemann tensor components are the same as in (\ref{Riem}) (with $A$
  replaced by $-A$) while for the metric (\ref{KS-tau}) they are
  (the dot denotes $d/d\tau$)
\bear                                                        \label{Riem-t}
      K_1 \eql \frac{\ddot a}{a}, \cm K_2 = \frac{\dot a \dot r}{ar},
\nn
      K_3 \eql \frac{\ddot r}{r}, \cm K_4 = \frac{1 + \dot r{}^2}{r^2}.
\ear

  The tidal forces acting on bodies at rest in the metric (\ref{KS-u}) or
  (\ref{KS-tau}) (comoving observers) are characterized by the same quantity
  (\ref{Z}), as before, while forces acting near a candidate horizon
  ($A \equiv a^2 \to 0$) on a noncomoving geodesic observer are proportional
  to $\tZ = Z/A$.

  It is easy to see that the components (\ref{Riem}) or (\ref{Riem-t})
  coincide with the corresponding tetrad components of the Riemann tensor
  and behave as scalars under radial or time coordinate changes.
  The quantities $Z$ and $\tZ$ also possess this property.

  Since the tensor $R\MN{}_{\rho\sigma}$ for the metrics under consideration
  is pairwise diagonal, the Kretschmann scalar ${\cal K}$ is a sum of
  squares,
\beq                                                         \label{Kr}
    {\cal K}= R\MN{}_{\rho\sigma} R^{\rho\sigma}{}\mn
        = 4K_1^2 + 8K_2^2 + 8K_3^2 + 4K_4^2,
\eeq
  and it is clear that all algebraic curvature invariants are finite (i.e.,
  a curvature singularity is absent) at a given point if and only if all the
  components (\ref{Riem}) or (\ref{Riem-t}) are finite.

%%%%%%%%%%%%%%%%%%%%%%%%%%%%%%%%%%%%%%%%%%
\section {Extension across a horizon}

  To study the extensibility of the metric beyond the surfaces that have
  been previously termed naked or truly naked horizons, it proves helpful to
  write the metric in the form (\ref{ds-u})or (\ref{KS-u}) because the
  coordinate $u$ is distinguished by the following properties
  \cite{vac1,cold}:
\begin{description}
\item[(i)]
  it always takes a finite value $u = u_h$ at a Killing horizon $A(u) =0$
  across which an extension is possible;
\item[(ii)]
  near a horizon, the increment $u - u_h$ is a multiple (with a nonzero
  constant factor) of the corresponding increments of manifestly
  well-behaved Kruskal-type null coordinates, used for analytic extension
  of the metric across the horizon.\footnote
    {Transitions through Killing horizons leading to a full space-time
    description and the corresponding Carter-Penrose diagrams have been
    considered in a general form in Refs.\,\cite{walk70,br79,strobl,kat};
    see also a detailed analytic treatment of the special but physically
        relevant case of the Reissner-Nordstr\"om extremal horizon in
    \cite{liber}.  }
  Therefore, with this coordinate, the geometry can be considered jointly on
  both sides of a horizon in terms of a formally static metric (hence the
  name ``quasiglobal'').
\end{description}

  To make the presentation complete, let us briefly prove items (i) and (ii)
  using, for certainty, the static metric (\ref{ds-u}).

  To begin with, the geodesic equations for (\ref{ds-u}) have
  integrals of the form
\beq                                                      \label{geo1}
     dt/d\lambda = E/A(u), \cm  d\phi/d\lambda = L/r^2(u),
\eeq
  and, combined with the normalization condition $u_\mu u^\mu = k$ for the
  tangent vector $u^\mu$ along the geodesic, they lead to
\beq                                                      \label{geo2}
     \biggl(\frac{du}{d \lambda}\biggr)^2
                  + A(u) \biggl(k + \frac{L^2}{r^2(u)}\biggr) = E^2.
\eeq
  Here, $\lambda$ is the canonical parameter, $k=1, \ 0,\ -1$ for timelike,
  null and spacelike geodesics, respectively; $L$ and $E\geq 0$ are constants
  characterizing the conserved angular momentum and energy of particles
  moving along the geodesics; without loss of generality, we consider curves
  in the equatorial plane $\theta = \pi/2$ of our coordinate system.

  As is clear from (\ref{geo2}), if $A \to 0$ as $u\to u_h$ (i.e., near a
  candidate horizon), then in its neighborhood
\beq
       d\lambda \approx  du/E,
\eeq
  unless $E=0$. Thus the $u$ coordinate behaves like a canonical parameter
  near $u = u_h$ for all geodesics except those with $E=0$. In particular,
  $\lambda \to \infty$ if $u_h = \infty$; a particle moving along a timelike
  geodesic reaches the surface $u = u_h$ at finite proper time if and only
  if $u_h$ is finite. The exceptional case $E=0$ corresponds to purely
  spatial curves, $t =\const$, see (\ref{geo1}), and then, according to
  (\ref{geo2}), $u \to \infty$ where $A\to 0$ again leads to $\lambda\to
  \infty$.

  We conclude that if $u \to \infty$ where $A\to 0$, the space-time is
  geodesically complete and no continuation is required. Such cases can be
  termed ``remote horizons'' and can be of interest by themselves but are
  irrelevant to the extension problem we consider here. Item (i) is proved.

  Let us now consider $|u_h| < \infty$ and introduce the so-called
  ``tortoise coordinate'' $x$ for the metric (\ref{ds-u}) by the relation
\beq
    dx = \pm du/A(u),                                  \label{x} %7
\eeq
  so that $A dt^2 - du^2/A = A(dt^2 - dx^2)$.
  Suppose, without loss of generality, $u > u_h =0$. Suppose, further, that
  in a finite neighborhood of $u=0$
\beq                                                                %8
    A(u) = u^n F(u), \cm n = \const,                    \label{A_h}
\eeq
  where $F(u)$ is a sufficiently smooth function such that $F(0)$ is finite.
  We should require either $n = 1$ or $n \geq 2$ to avoid a curvature
  singularity at $u=0$ [see the first line in (\ref{Riem})]. Then, by
  (\ref{x}), $x\to \pm\infty$ as $u\to 0$; we choose, for certainty,
  $-\infty$.

  To cross the surface $u=0$, we introduce, as usual, the null coordinates
\beq
    V = t+x, \cm W = t-x,                              \label{VW} %9
\eeq
  The limit $V \to -\infty$ corresponds, at fixed finite $W$, to a past
  horizon, and $W\to \infty$ at fixed finite $V$ to a future horizon.
  Introducing the new null coordinates $v = v(V)$ and $w = w(W)$
  and properly choosing these functions, we can compensate the zero of $A$
  in the expression $A(dt^2-dx^2) = A dV dW$ at $V =-\infty$ or $W =\infty$.

  Consider the future horizon $W = +\infty$ at fixed finite $V$. Then,
  a finite value of the metric coefficient $g_{Vw}$ at the horizon is
  obtained under the condition that in its neighborhood
\beq
    dw/dW \sim A.                                       \label{A1} %11
\eeq
  On the other hand, according to (\ref{x}), $A(u) = du/dx$, while, by
  (\ref{VW}), $x \approx -2W$ near the future horizon, whence it follows
\beq
        A(u) \approx -2 du/dW.                              \label{A2} %12
\eeq
  Comparing (\ref{A1}) and (\ref{A2}), we see that
\bearr
     du\sim dw\ \ \ \mbox{at the future horizon},
\nnn                                                        \label{Krus} %13
     du\sim dv\ \ \ \mbox{at the past horizon},
\ear
  where the second line is obtained in the same way as the first one if we
  choose $V(v)$ to make the quantity $g_{vW}$ finite. This proves item (ii).

  The coordinates $(v,w)$ are null Kruskal-like coordinates suitable for
  obtaining a coordinate system valid on both sides of a horizon. Like
  $u$, they are finite at the horizon (say, $v=0$ and $w=0$), and, to admit
  continuation across it, the metric should be analytic in terms of $v$
  and $w$ at $v=0$ and $w=0$, respectively, and consequently analytic
  in terms of $u$ at $u = u_h$.

  Thus the functions $A(u)$ and $r^2(u)$ in (\ref{ds-u}) must be analytic,
  and a horizon corresponds to a regular zero of $A(u)$, i.e., $A(u) \sim
  (u-u_h)^n$, where $n \in \N$ is the order of the horizon.  Or, as a
  minimal requirement, they should belong to a class $C^s$ with $s\geq 2$.
  In the latter case, a continuation across the horizon will have the order
  of smoothness $s$, but then discontinuities in derivatives of orders
  higher than $s$ should be somehow physically justified, e.g., by
  boundaries in matter distributions.

  More specifically, assuming (\ref{A_h}), we obtain
\beq                                                               %% 14
          u \sim \vars{\e^{xF(0)}       \quad & (n=1),\\       \label{ux}
                       |x|^{-1/(n-1)}   \quad & (n > 1).}
\eeq
  as $x\to -\infty$, and accordingly, at the future horizon ($W\to\infty$)
\beq
      w \sim \vars{ \e^{-WF(0)/2} \quad  & (n=1),\\     \label{wW} %%15
                     W^{-1/(n-1)} \quad  & (n > 1)}
\eeq
  [recall that, by (\ref{VW}), $x \approx -W/2$ as $W\to \infty$]. Replacing
  $w\to v$ and $W \to |V|$, we obtain similar relations valid at the past
  horizon ($V\to -\infty$).

  Crossing the horizon corresponds to a smooth transition of the coordinate
  $v$ or $w$ from positive to negative values. This clearly explains why
  such transitions are impossible if analyticity is lacking: if, for
  instance, $A\sim u^n$ with fractional $n$, the transformed metric will
  inevitably contain fractional powers of $v$ and $w$, which are either
  meaningless (if $n$ is irrational) or at least not uniquely defined at
  negative $v$ and $w$.

%%%%%%%%%%%%%%%%%%%%%%%%%%%%%%%%%%%%%%
\section {Static metrics: horizons vs. singularities}

  We have seen that, to be regular at a candidate horizon (i.e., a sphere
  where $g_{tt} = 0$) and analytically extensible beyond it, the metric
  should be analytic in terms of the quasiglobal coordinate $u$. Thus, at
  such a horizon,
\begin{description}
\item[(a)]
    $u=u_h$ should be finite, and we can put without loss of generality
    $u_h =0$ (and consider $u > 0$ as the static region);
\item[(b)]
    the value $u=0$ should be a regular zero of the function $A(u)$
    i.e., $A(u) \sim u^n$, $n \in \N$ being the order of the horizon;
\item[(c)]
    the function $r(u)$ should behave near $u=0$ as
\beq                                                        \label{r(u)}
        r(u) = r_h + r_k u^k + o(u^k),
\eeq
  where $k\in \N$ is the order of the second nonvanishing term of the
  Taylor series.
\end{description}

  To see how the metric taken in the general form (\ref{ds}) behaves at a
  horizon, let us identify it with (\ref{ds-u}) term by term, so that
\beq
     \e^{2\gamma(R)} = A(u), \cm                             \label{iden-R}
     \e^{\alpha(R)}dR = A^{-1/2}du, \cm
     r(R) = r(u),
\eeq
  and use the fact that not only $g_{tt}$ and $g_{\theta\theta}$ are
  reparametrization-invariant (behave like scalars when we change the
  coordinate $R$) but also the expressions $\e^{-\alpha}\gamma_R$ and
  $\e^{-\alpha}r_R$ where the subscript $R$ stands for $d/dR$.
  Taking into account that $A = \e^{2\gamma} \sim u^n$, we obtain:
\bearr                                                        \label{gam'}
     \e^{-\alpha}\gamma_R \sim u^{(n-2)/2} \sim \e^{(n-2)\gamma/n},
\\ \lal                                                       \label{r'}
     \e^{-\alpha}r_R \sim u^{(n+2k-2)/2} \sim \e^{(n+2k-2)\gamma/n}.
\ear
  These are necessary conditions of a regular horizon at a value of $R$
  where $\e^{\gamma} =0$ in terms of the general metric (\ref{ds}).

  Let us discuss in more detail the most popular coordinate choice,
  $R=r$, such that
\beq                                                         \label{ds-r}
     ds^2 = \e^{2\gamma(r)} dt^2 - \e^{2\alpha(r)}dr^2 - r^2 d\Omega^2,
\eeq
  and the transformation (\ref{iden-R}) now reads
\beq
     \e^{2\gamma(r)} = A(u), \cm                             \label{iden-r}
     \e^{\alpha(r)}dr = A^{-1/2}du, \cm  r = r(u).
\eeq

  Following \cite{Z}, we suppose that near $r=r_h$
\beq                                                         \label{p,q}
    \e^{2\gamma} \sim (r-r_h)^q, \cm \e^{2\alpha}\sim (r-r_h)^{-p},
\eeq
  where $q >0$ and, as follows from the curvature regularity requirement [see
  (\ref{Riem})], it should be
\beq                                                        \label{reg-r}
       {\rm either}\ \ p\geq 2\ \ {\rm or}\ \ 2> p\geq 1, \ \ p + q = 2.
\eeq

  The condition that $u$ should be finite at $r=r_h$ leads to
\beq
        q > p - 2.                                  	    \label{u-fin}
\eeq
  Furthermore, \eq (\ref{gam'}) selects a sequence of lines in the
  ($p,q$) plane, intersecting at the point $O$: $(2,0)$ (see Fig.\,1),
\beq
      L_n: \cm q(n-2) = n (p-2).                             \label{pq1}
\eeq
  Finally, \eq (\ref{r'}) leads to the condition
\beq                                                         \label{pq2}
      np = q(n+2k-2).
\eeq
  From \eqs (\ref{pq1}) and (\ref{pq2}) we see that the regularity
  conditions select a discrete set of points in the ($p,q$) plane,
  parametrized by two positive integers $n$ and $k$:
\beq
      p = 2 + (n-2)/k, \cm q = n/k.                          \label{pq3}
\eeq

  In the generic case $k=1$, we obtain the important points $H_n:\ p = q =
  n$. Then, at least in a neighborhood of $r=r_h$, $r(u)$ is a linear
  function.

  The following remark is in order. Actually, \eq (\ref{pq3}) expresses $p$
  and $q$ defined in (\ref{p,q}) in terms of $n$ (such that $A \sim u^n$)
  and $k$ defined in (\ref{r(u)}) for any $n$ and $k$, not necessarily
  integers. The inverse expressions, valid for $q > p-2$, are
\beq
    n = 2q/(q-p+2), \cm k = 2/(q-p+2).                     \label{nk}
\eeq
  \eqs (\ref{pq3}) and (\ref{nk}) are useful in dealing with examples to be
  discussed in Sec.\,V.

  If we weaken the above item (c) and only require that the quantities
  $r'$ and $r''$, entering into the Einstein equations, should be finite,
  then, instead of (\ref{pq2}), we obtain that either $p = q$ (again giving
  the points $H_n$) or $q \leq p-1$. Thus, in the $(p,q)$ plane, horizons
  correspond to the sequence of points $H_n$ plus a sequence of segments of
  the lines $L_n$ [\eq (\ref{pq1})] belonging to the band (see Fig.\,1)
\beq
      p-2 < q \leq p-1.                                       \label{band}
\eeq

\begin{figure} \centering
    \epsfxsize=125mm\epsfbox{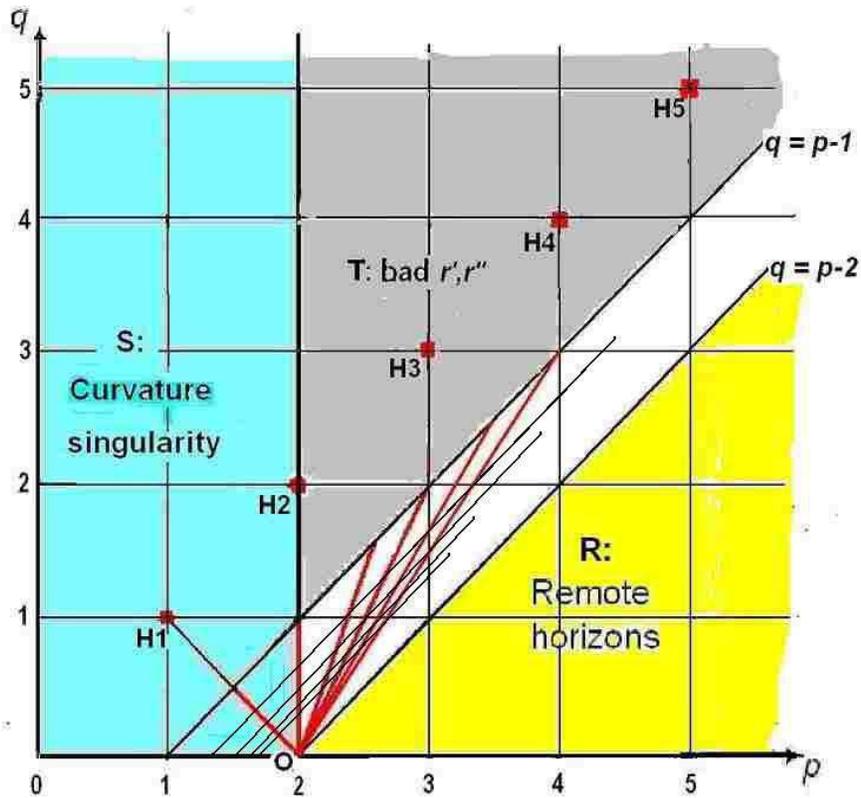}
%%\bigskip
\caption{\protect\small
    Properties of the candidate horizon $r=r_h$ depending on $p$ and $q$.
    Horizons admitting extensions correspond to the points H1,\ H2,...
    ($p = q\in \N$) and other points (\ref{pq3}), some of which (with
    $k= 2, \ldots, 6$) are shown by intersections of the beams
    (\ref{pq1}) originating at point O with the line $q=p-1$ and with
    the thin lines parallel to it.  Under the weakened condition on
    $r(u)$, namely, that $r'$ and $r''$ should be finite, admissible are
    whole segments of the beams (\ref{pq1}), belonging to the band
    (\ref{band}). In region S ($p < 2$), outside the segment OH1,
    $r=r_h$ is a curvature  singularity.  Outside the other beams
    (\ref{pq1}), an extension of the metric beyond the sphere $r = r_h$
    is impeded by fractional powers in $A(u)$ and, in region T, by
    infinities in $r'$ (in case $q > p$) and $r''$. In region R,
    including the line $q = p-2$, $r=r_h$ is a remote horizon.}
\medskip\hrule
\end{figure}

  Now we are ready to compare the above results with the classification of
  horizons \cite{ZP,Z} as ``usual'', ``naked'' and ``truly naked'' in
  terms of the metric (\ref{ds-r}) under the assumption (\ref{p,q}). A
  criterion for distinguishing them is \cite{Z}
\beq                                                 \label{crit}
\renewcommand{\arraystretch}{1.1}
\begin{array}{ll}
    {\rm usual:} \quad      & \tZ (r_h) =0, \\
    {\rm naked:} \quad      & \tZ (r_h) \ {\rm finite},\\
    {\rm truly\ naked:}  \quad  & \tZ (r_h) = \infty,
\end{array}
\eeq
  where the quantity $\tZ$, characterizing the magnitude of tidal forces at
  $r=r_h$ in a freely falling reference frame, is given by [see (\ref{tZ})]
\beq
     \tZ = \const \cdot (q-p)(r-r_h)^{p-1-q}.              \label{tZ1}
\eeq
  Its counterpart in the static reference frame is $Z = \const\cdot
  (q-p)(r-r_h)^{p-1}$, which is zero at all candidate horizons.

  We would like to stress that only transversal components of the curvature
  tensor are relevant to our discussion. The reason is that
  the $R^{tr}{}_{tr}$ component of the Riemann tensor, which characterizes
  tidal forces in the radial direction, does not change under radial boosts.
  So if it is finite, it is irrelevant to the issue of singularities while
  if it diverges, we evidently deal with a curvature singularity for
  which the Kretschamnn scalar diverges.

  The results of such a comparison are collected in Table I.
  The table mentions one more quantity, $Q$, that distinguishes usual and
  naked horizons in the case $p=q$ \cite{Z},
\beq
      Q = 2(\alpha_r + \gamma_r),                 \label{Q}
\eeq
  which depends on further details of the behavior of $\alpha$ and
  $\gamma$ as compared with (\ref{p,q}).

\begin{table} %% I
  \caption{Horizon types according to Refs.\,\cite{ZP,Z} and the results
       of the present analysis. Curvature ($C^0$) singularities are
       not included.}

\medskip
\centering
\renewcommand{\arraystretch}{1.15}
\tabcolsep 4pt
\begin{tabular}{|c|l|l|l|}
\hline
  No. &  $p,\ q$    & type by (\ref{crit}) & present analysis \\
\hhline{|=|=|=|=|}
  1   & $p = q = 1$ & usual or naked$^*$ & regular, $n=1$    \\
\hline
  2   & $1 < p < \fract 32$, $q=2-p$
            & truly naked        & singular         \\
\hline
  3   & $p=\fract 32,\ q=\fract 12$
            & naked              & regular, $n=1$    \\
\hline
  4   & $\fract 32 < p < 2$, $q=2-p$
            & usual              & regular, $n=1$    \\
\hline
  5   & $p \geq 2,\ q > p$
            & truly naked        & singular         \\
\hline
  6   & $p \geq 2,\ q=p$
            & usual or naked$^*$ &
                        regular if $p=q=n$, otherwise singular \\
\hline
  7   & $p \geq 2,\ p-1 < q < p$
            & truly naked        & singular         \\
\hline
  8   & $p \geq 2,\ q = p-1$
            & naked              &
                    regular if $p=1+n/2$, otherwise singular \\
\hline
  9   & $p \geq 2,\ p-2 < q < p-1$
            & usual              &
              regular if (\ref{pq1}) holds, $n\in \N$, otherwise singular \\
\hline
  10  & $p \geq 2,\ q \leq p-2$
            & usual              & remote horizon   \\
\hline
\end{tabular}
\medskip

  $^*$ The horizon is usual if $Q=0$ and naked if $Q\ne 0$ at $r=r_h$ with
  $Q$ defined in \eq (\ref{Q}).

\end{table}

  In the fourth column, the word `regular' means that the above criteria (a),
  (b) are fulfilled, $r'$ and $r''$ are finite, and an (at least $C^2$)
  extension is possible; though, on the segments (\ref{pq1}) (\ref{band})
  (lines 4, 9 of the table), this extension can be analytic only at the
  countable number of points (\ref{pq3}). The term `singular' means
  impossibility of an extension, though all algebraic curvature invariants
  are finite.

  From the viewpoint of a general classification of space-time singularities
  (see, e.g., \cite{e-sch77}), what we usually call a curvature singularity
  [a set of (limiting) space-time points with an infinite value (or
  discontinuity) of at least one algebraic invariant of the Riemann tensor]
  is called a $C^0$ curvature singularity. A more general notion is that of
  a $C^m$ curvature singularities, related to a discontinuity in some of the
  tetrad components of the $m$-th covariant derivatives of the Riemann
  tensor. (In many cases, but not always, at such singularities, some
  invariants involving derivatives of the Riemann tensor are discontinuous.)
  The singularities discussed here are apparently related to such
  higher-order curvature singularities, but this relation is quite
  nontrivial, and its exact formulation requires a separate study which is
  beyond the scope of this paper.

  Table I shows that all TNHs and (somewhat unexpectedly) many of the
  ``simply'' naked horizons and even some ``usual'' ones are actually
  space-time singularities in the sense that geodesics terminate there at
  finite values of the canonical parameter.

  One can remark that analytic properties of the metric near a horizon are
  often discussed in terms of the two-dimensional ($t,R$) submanifold,
  assuming the angular part of the metric to be irrelevant. We have seen
  that, on the contrary, it is the behavior of the derivatives $r'$ and
  $r''$ that makes singular many candidate horizons looking quite regular in
  the ($t,R$) submanifold.

\section{Examples}

\subsection{Naked and ``potentially naked'' horizons}

  At event horizons of naked \bhs\ according to \cite{Z, ZP}, the quantity
  $\tZ$ [see (\ref{tZ}) or (\ref{tZ1})] is finite. To be naked according to
  \cite{HR1, HR2}, such a horizon must exhibit large values of $\tZ$ on
  the Planck scale while the curvature invariants remain small on the same
  scale. Let us discuss the relationship of these notions using as an
  example the dilaton \bh\ solution \cite{bsh77,GM,GHS92}, discussed in
  \cite{HR1}, with the metric (\ref{ds-u}) such that
\beq
    A = \frac{(u-u_+)(u-u_-)}{r^2},\cm                     \label{dil}
    r = u \biggl(1 - \frac{u_{-}}{u}\biggr)^b, \cm 0 < b < 1,
\eeq
  with $b = \const$. There is a horizon at $u = u_+ > u_-$ and a singularity
  at $u=u_-$. The extremal limit corresponds to $u_+ = u_-$, and in this
  limit the horizon area shrinks, $r_+ = r(u_+) \to 0$, leading to a
  singularity.

  Then, simple calculations give
\beq                                                        \label{Z-dil}
    Z=-\frac{r''}{r}A, \cm
    \tZ \sim \frac{Z}{A} = -\frac{r''}{r}
        = -\frac{u_- b(b-1)}{u^3}\biggl(1-\frac{u_-}{u}\biggr)^{-2}.
\eeq
  We see that $\tZ(u_+)$ is finite, so that the horizon is naked according
  to \cite{Z, ZP}.

  The curvature components in the static frame are of the order $1/r_+^2$.
%%%%%  \sim \frac{1}{u^{2}}(1-\frac{u_{-}}{u})^{-b}$.
  Introducing the parameter $\eps = 1- r_-/r_+$, we obtain that near the
  horizon $\tZ \sim u_+^{-2}\eps^{-2}$ and $r_+^{-2}\sim u_+^{-2}\eps^{-b}$.
  In the limit $\eps \to 0$, both $\tZ$ and the curvature invariants become
  infinite as $u \to u_+=u_-$ since $ u = u_-$ is a singularity.
  If, however, one chooses an intermediate value of $\eps$ such that
\beq
    \left(\frac{l_{Pl}}{u_{+}}\right)^{2/b}\ll \eps \ll 1, \label{e}
\eeq
  so that $u_+ \gg l_{Pl}$, where $l_{Pl}$ is the Planck length, we arrive
  at the situation that the tidal forces are large but the curvature
  components in the static frame (and hence the curvature invariants) remain
  small on the Planck scale \cite{HR1}.

  Thus the naked behavior of this black hole is due to proximity to the
  singularity. As to the general case (\ref{dil}), (\ref{Z-dil}), its
  naked nature according to (\ref{crit}) \cite{ZP, Z} only indicates that
  this solution can really exhibit a naked behavior at some values of its
  parameters. In this sense, as was said in the Introduction, such horizons
  (and \bhs) may be called ``potentially naked'' as opposed to naked ones
  according to Horowitz and Ross and truly naked ones which are always
  singular as follows from our present results.

\subsection{Matter with an arbitrary equation of state}

  Consider again the metric (\ref{ds-u}) and the Einstein equations for it
  with the stress-energy tensor $T\mN = T\mN{}_{(m)} + T\mN\vac$ where
\beq                                                           \label{T_mat}
     T_{\mu (m)}^{\nu } = \diag (\rho,\ -p_r, -p_\bot,\ -p_\bot).
\eeq
  is the contribution of some kind of matter (e.g., a perfect fluid) and
\beq                                                          \label{T_vac}
     T\mN\vac = \diag (\rho\vac,\ \rho\vac,\ -p_\bot\vac,\  -p_\bot\vac).
\eeq
  is that of a so-called vacuum fluid, distinguished by the property
  $T_0^0 = T_1^1$ \cite{dym}. Particular cases of such ``vacua'' are a
  cosmological constant ($p_\bot\vac = -\rho\vac = -\Lambda/(8\pi)$),
  electric or magnetic fields in the radial direction ($p_\bot\vac =
  \rho\vac$) and other forms which may be specified, e.g., by $\rho\vac$ as
  a function of $r$ \cite{dym, br-NED, eliz1, eliz2}.

  Let us choose, as independent components of the Einstein equations, the
  difference ${0\choose 0} - {1\choose 1}$ and the ${1\choose 1}$ equation
  which read
\bear
    2A\frac{r''}{r} \eql -8\pi (\rho + p_r),        \label{0-1}
\\
     \frac{1}{r^2}(-1 + A'rr'+ Ar'{}^2)\eql 8\pi(p_r -\rho\vac). \label{11}
\ear
  In addition, assuming no interaction between matter and vacuum, we can
  write the conservation law for matter in the form
\beq                                                        \label{cons}
    p'_r+\frac{2r'}{r} (p_r - p_\bot) + \frac{A'}{2A}(\rho +p_r)=0.
\eeq
  and similarly for vacuum.

  Let us also assume that the matter obeys the linear equation of state
  (at least approximately, near the horizon)
\beq
    p_r \approx w\rho, \ \ \ w = \const.                    \label{w}
\eeq
  Then, a further analysis can be carried out along the lines of \cite{bz2},
  with the only difference that now we do not restrict ourselves to only
  regular horizons and seek any solutions with $A\to 0$ as $u\to u_h$ where
  $u_h \ne \infty$. In particular, we do not require that $r'$ and $r''$
  should be finite as $u \to u_h$. Omitting the details, we present the
  results.

  We obtain that if there is no vacuum ($T\mN\vac =0$), then the only
  possible solutions correspond to a simple horizon, such that $A(u) \sim
  u-u_h$, and, provided $w = -1/(1+2k)$ where $k$ is a positive integer,
  we obtain regular solutions in full agreement with \cite{bz2}. Solutions
  with TNHs are not obtained. The value $w = -1$ is excluded since it
  corresponds to a vacuum fluid, contrary to what was assumed.

  If, however, we consider a mixture of the two kinds of matter described by
  (\ref{T_mat}) and (\ref{T_vac}), there appear solutions containing
  matter with $0 > w > -1$ and $\rho \sim A^{-(w+1)/(2w)}$. Furthermore, if
  we turn to the curvature coordinates and assume that the metric
  coefficients behave according to (\ref{p,q}), we obtain, for $p\ne q$, the
  following relation:
\beq
    w = - q/(q + 2p -2),                                   \label{wpq}
\eeq
  and it appears that $\rho >0$ for $q > p$ and $\rho < 0$ for $q < p$.
  Thus any $p$ and $q$ satisfying the condition (\ref{reg-r}) are admissible,
  except for those with $p = q$. In the latter case, solutions can also
  exist, with $w$ satisfying the requirement $w > -p/(3p-2)$.

  All kinds of solutions mentioned in Table I are possible, and the values
  of $w$ cover the whole range from 0 to $-1$. This is related to the
  underdetermined nature of the system since the function $\rho\vac(u)$
  remains arbitrary.

  If we put $\rho\vac = \Lambda/(8\pi) = \const$, thus specifying the vacuum
  as a cosmological constant, the Einstein equations relate the exponents
  $p$ and $q$ characterizing the metric to the matter parameter $w$.
  Namely, we have either (i) $p =q =1$ (a simple regular horizon) and
  $w = -1/(1+2k)$ (as described in \cite{bz2}) or (ii)
\beq
    p=2, \cm q = -2w/(w+1), \cm w \ne -1/2.               \label{p/Lam}
\eeq
  The parameter $w$ can take any value in the range $(-1,\ 0)$ except
  $-1/2$. The configurations correspond to lines 5, 7, 8, 9 of Table I.

\subsection {Scalar field with a potential}

  Consider a real, minimally coupled scalar field $\phi$ with the Lagrangian
  $L_s = (16\pi)^{-1} [\ep g\MN \d_\mu\phi \d_\nu \phi - 2V(\phi)]$ as a
  source of gravity. Here, $V(\phi)$ is a potential while $\ep =1$
  corresponds to normal fields with a correct sign of the kinetic energy and
  $\ep = -1$ to the so-called phantom fields, often discussed in the context
  of the dark energy problem.

  Then \ssph\ configurations with the metric (\ref{ds-u}) and $\phi =
  \phi(u)$ obey the equations
\bear
     (Ar^2 \phi')' \eql  \eps r^2 dV/d\phi,                \label{phi''}
\\
              (A'r^2)' \eql - 2r^2 V,                           \label{00}
\\
              2 r''/r \eql -\eps{\phi'}^2 ,                     \label{01}
\\
         A (r^2)'' - r^2 A'' \eql 2.                            \label{02}
\ear
  \eq (\ref{phi''}) follows from (\ref{00})--(\ref{02}), which, given a
  potential $V(\phi)$, form a determined set of equations for the unknowns
  $r(u),\ A(u),\ \phi(u)$. \eq(\ref{02}) is once integrated giving
\bear
    \biggl(\frac{A}{r^2}\biggr)'
                      = \frac{2(u_0 -u)}{r^4},          \label{A'}
\ear
  where $u_0$ is an integration constant.

  Let us assume that $u=0$ is a candidate horizon, near which
\beq                                                           \label{A,r}
    A \approx A_n u^n(1 + o(1)), \cm r = r_h + r_k u^k + o(u^k),
\eeq
  where $n >0,\ A_n > 0,\ r_h >0,\ r_k$ are constants. According to Sec.\,IV,
  at a regular horizon, $n$ and $k$ should be positive integers. In
  particular, $|r''| < \infty$, and \eq (\ref{01}) yields a finite value of
  $\phi'$. As a result, one can arrive at different \bh\ solutions with
  particular potentials (see, e.g., \cite{mann,vac1,vac2} and references
  therein), including globally regular solutions for phantom fields (the
  so-called black universes \cite{bu12}).

  Our interest here is, however, in the existence of solutions containing
  TNHs, let us therefore assume $k\not\in \N$. Then \eq (\ref{01}) in the
  leading order of magnitude gives
\beq                                                          \label{phi'}
    \phi' \approx \pm \phi_1 u^{k/2-1}, \cm
                      \phi_1 := \sqrt{-2\ep k(k-1)r_k/r_h},
\eeq
  so that, evidently, we must have $(k-1)\ep r_k < 0$.
  Choosing the upper sign and integrating, we have
\beq                                                         \label{phi}
     \phi \approx \phi_0 + \frac{2\phi_1}{k} u^{k/2}, \cm \phi_0=\const.
\eeq

  On the other hand, \eq (\ref{A'}) in the leading order of magnitude gives
  different results for $u_0 =0$ and $u_0\ne 0$, namely,
\bearr                                                       \label{A1,2}
     n=1, \cm A_1 r_h^2 = 2u_0 \cm {\rm for}\ \ u_0 \ne 0,
\nnn
     n=2, \cm A_2 r_h^2 = -1 \ \ \cm {\rm for}\ \ u_0 =0.
\ear
  The second case is excluded due to the requirement $A_2 > 0$. So we are
  left with the first line in (\ref{A1,2}), and the main point is that
  $A(u)$ is automatically well-behaved and corresponds to a simple horizon
  ($n=1$) irrespective of the value of $k$.

  Substituting the expressions for $r$ and $A$ into \eq (\ref{00}), we find
  that $V$ tends to infinity as $u\to 0$ (thus leading to a curvature
  singularity) if $k<1$ and to a finite limit if $k > 1$. Thus we must have
  $k >1 $ in \eq (\ref{A,r}).

  An asymptotic form of $V(\phi)$ as $\phi\to\phi_0$ is easily obtained
  using \eq(\ref{phi''}) which yields $dV/d\phi$. As a result, we have
\beq                                                            \label{V}
       V(\phi)\approx  V_0 + \const\cdot (\phi-\phi_0)^{2(k-1)/k}.
\eeq

  Thus there exist solutions in the form of singular horizons for potentials
  having the asymptotic form (\ref{V}), i.e., at values $\phi=\phi_0$, if
  any, approached with fractional powers of $\phi-\phi_0$ (recall that we
  have been considering $k \not\in \N$).

  Passing over to the curvature coordinates using (\ref{iden-r}), we see
  that the above asymptotic solution is described by the conditions
  (\ref{p,q}) where $p$ and $q$ are expressed in terms of $k$ according to
  (\ref{pq3}) with $n=1$, i.e., $p = 2 - 1/k$, $q = 1/k$. In other words,
  these solutions reside on the segment OH1 in Fig.\,1 and correspond to
  lines 2,3,4 in Table I.

\subsection {$(2+1)$-dimensional analogue: an exact solution}

  We have given examples of TNHs in some solutions of general relativity by
  analyzing the near-horizon geometry. Unlike that, in $(2+1)$-dimensional
  general relativity it is possible to present an exact solution with a
  horizon, where the source of gravity is a perfect fluid with a linear
  equation of state, such that
\beq                                                             \label{T3}
    T\mN = \diag(\rho,\ -p,\ -p), \cm  p = w\rho, \ \ w=\const.
\eeq
  This solution can be extracted from the results of Ref.\,\cite{ext},
  a study aimed at finding all possible static analogues of the
  Bertotti-Robinson space-time without requiring spherical symmetry. A
  $(2+1)$-dimensional section of one of the classes of space-times obtained
  there has a circularly symmetric metric similar to 4D spherically
  symmetric metrics with a TNH. The metric has the form
\beq                                \label{ds_3}
    ds^2 = \e^{2\gamma} dt^2 - \e^{2\alpha} dr^2 -r^2 d\phi^2,
\eeq
  where
\beq                                                            \label{g_3}
    \e^{2\alpha} = \left(\frac{r_0^2 - r^2}{r_0^2}\right)^{-2/(1-w)},
    \cm
        \e^{2\gamma} = \const\cdot (r_0^2-r^2)^{-2w/(1-w)},
\eeq
  and the density is
\beq                                                            \label{rho_3}
    \rho = \frac{(r_0^2 - r^2)^\eta}{4\pi (1-w)r_0^{4/(1-w)}},
           \cm \eta = \frac{1+w}{1-w}.
\eeq
  A candidate horizon, at which $\e^{\gamma} =0$, is $r=r_0$ if $w <0$ or
  $w > 1$. In its neighborhood,
\beq                                                         \label{Z3}
    \tZ \sim \e^{-2\gamma} Z \sim \e^{-2\gamma} \rho
                \sim (r_0^2 - r^2)^{\eps},\cm
    \eps = \frac{1+3w}{1-w}.
\eeq
  \eq (\ref{Z3}) does not work for $w=-1$ corresponding to a cosmological
  constant and 3D de Sitter metric, in which case $Z = \tZ = 0$.

  The density is finite or zero at $r=r_0$ if $\eta \geq 0$, and it can be
  verified that the Kretschmann scalar is finite under the same condition.
  Thus a curvature singularity at $r=r_0$ is absent in the whole range $-1
  \leq w < 1$, but possible horizons for $w > 1$ are excluded. On the other
  hand, if $\eps < 0$, then $\tZ$ diverges, so that the tidal forces in the
  freely falling frame are infinite, and for
\beq
       -1 < w < -1/3                                           \label{w_3}
\eeq
  we have a solution with a TNH. Recall for comparison that in 4 dimensions
  (Sec. V.B) TNHs appeared in solutions with a similar source in the
  range $-1 < w < 0$ and only in the presence of a vacuum fluid.

\begin{table}  %% II
  \caption{Horizon types in the solution (\ref{ds_3})--(\ref{rho_3}) in the
   range $-1 \leq w < 1$.}

\medskip
\centering
\renewcommand{\arraystretch}{1.15}
\tabcolsep 8pt
\begin{tabular}{|c|l|l|l|}
\hline
  No. &  $w$    & horizon type by (\ref{crit}) & present analysis \\
\hhline{|=|=|=|=|}
  1   & $w \in (0,\,1)$  & \multicolumn{2}{|l|}{\inch no horizon}    \\
\hline
  2   & $w \in (-1/3,\,0)$
            & usual              & regular if (\ref{w3}) holds,
                           otherwise singular   \\
\hline
  3   & $w = -1/3$
            & naked              & regular, $n=1,\ k=2$  \\
\hline
  4   & $w \in (-1,\,-1/3)$
            & truly naked        & singular         \\
\hline
  5   & $w = -1$    & usual          & regular, $n=1,\ k=1$ \\
\hline
\end{tabular}
\medskip

\end{table}

  An analysis similar to that of Sec.\,IV leads to the results presented in
  Table II. The terminology is the same as in Table I: e.g., the word
  ``regular'' means that the metric can be extended beyond the horizon. It is
  of interest that a transition to the quasiglobal coordinate $u$ similar to
  (\ref{iden-r}) leads, for any $w < 0$, to $A \sim u$, i.e., to a
  first-order horizon ($n=1$), which is, however, not always regular due to
  the behavior of $r(u)$. To satisfy the regularity criterion (c) [see \eq
  (\ref{r(u)})], the parameter $w$ should be related to $k\in \N$ by
\beq
    w = -1/(2k-1).                                         \label{w3}
\eeq

  As in 4 dimensions, all TNHs turn out to be singular, and even those
  horizons that seem usual can be singular as well (see line 2 of Table II).

\section {Horizons in \KS\ cosmologies}

\subsection {Regular horizons}

  It is easy to reformulate the regular horizon conditions like (\ref{gam'})
  and (\ref{r'}) for a time-dependent homogeneous analogue of \ssph\
  space-times, i.e., KS cosmologies.

  A KS metric written in terms of an arbitrary time coordinate $t$ reads
\beq
    ds^2 = b^2(t) dt^2 - a^2(t) dx^2 - r^2(t) d\Omega^2.  \label{KS-gen}
\eeq
  with an arbitrary lapse function $b(t)$ and two scale factors $a(t)$ and
  $r(t)$. Identifying the metrics (\ref{KS-u}) and (\ref{KS-gen}) and
  applying the requirements (a), (b), (c) from Sec.\,IV to (\ref{KS-u})
  precisely as in static spherical symmetry, we arrive at the following
  necessary conditions of a regular (extensible) horizon at a value of $t$
  at which $a\to 0$:
\bearr
     \frac{1}{b}\frac{da}{dt}\sim u^{n-1} \sim a^{2(n-1)/n},  \label{a-dot}
\\ \lal                                                       \label{r-dot}
    \frac{1}{b}\frac{dr}{dt}\sim u^{n/2 + k - 1} \sim a^{(n+2k-2)/n}
\ear
  Here, as before, $n\in\N$ is the order of the horizon and $k\in \N$
  corresponds to \eq (\ref{r(u)}).

  One can also modify the results of Sec.\,IV for a cosmological analogue of
  the curvature coordinate, i.e., the scale factor $r(u)$ considered as
  a coordinate. It seems, however, more helpful to present detailed horizon
  regularity conditions for the metric (\ref{KS-tau}) written in terms of
  the proper time $t=\tau$, most widely used in cosmology as a natural time
  variable. Identifying (\ref{KS-u}) and (\ref{KS-tau}) term by term, we
  have
\beq
    d\tau =\pm \frac{du}{\sqrt{A(u)}}, \cm a^2(\tau) = A(u), \cm
    r(\tau) = r(u).                                         \label{iden-KS}
\eeq

  Applying again the requirements (a), (b), (c) from Sec.\,IV, we obtain
  the following properties of the metric (\ref{KS-tau}) at a regular horizon
  occurring at a value of $\tau$ where $a \to 0$ and $r\to r_h>0$.

\medskip\noi
{\bf 1.} A horizon may correspond to finite or infinite $\tau$. In the
  latter case, the integral $u = \int a(\tau) d\tau$ should still converge,
  otherwise we deal with a remote horizon in the absolute past or future,
  attained by all geodesics at infinite values of their canonical parameters.

\medskip\noi
{\bf 2.} A first-order horizon ($n=1$) occurs at finite $\tau = \tau_h$,
  near which
\beq
    a(\tau) \sim \tau-\tau_h, \cm
    r(\tau) \approx r_h + C (\tau-\tau_h)^{2k},                 \label{hor1}
\eeq
  where $r_h$ and $C$ are constants and $k\in \N$ is the exponent from the
  condition (\ref{r(u)}).

\medskip\noi
{\bf 3.} A second-order horizon ($n=2$) is characterized by
\beq                                                            \label{hor2}
     \tau \to \pm\infty, \cm a(\tau) \sim \e^{-m|\tau|}, \cm
               r(\tau) \approx r_h + C\e^{-km|\tau|}
\eeq
   with $m = \const > 0$.

\medskip\noi
{\bf 4.} At higher-order horizons ($n \geq 3$), we have
\beq                                                            \label{hor3}
     \tau \to \pm\infty, \cm a(\tau) \sim |\tau|^{-n/(n-2)}, \cm
               r(\tau) \approx r_h + C|\tau|^{-2k/(n-2)}.
\eeq

  Thus horizons of any order $n$ are possible in the past (or future)
  of a KS universe. As argued in \cite{bd07, bz1}, a matter source for such
  a behavior of the metric is probably pure vacuum since, among other kinds
  of matter, only that with some particular values of $w \leq -3$, which may
  be called ``deeply phantom'', is admissible \cite{bz1}. It has been
  concluded \cite{bd07, bz1} that normal matter could have been created
  later from vacuum along with isotropization.

  We also see that regular horizons of orders 2 and higher occur at infinite
  proper times of observers at rest in KS space-times. In other words, if
  the present cosmological evolution began at such a horizon, it has
  happened infinitely long ago. However, other geodesic paths, both timelike
  and null, cross such horizons at finite values of their canonical
  parameters (due to finite quasiglobal time $u$), and one can conclude that
  one-way communication with regions beyond such horizons is still possible
  (this phenomenon was discussed in detail for vacuum KS cosmologies in
  \cite{bdd03}). If we live in such a universe, we can in principle catch
  photons or massive particles coming from an infinitely remote past
  according to our own clocks.

  This situation is fundamentally different from that of a remote horizon in
  the past, where $u\to -\infty$, which is also possible but \cite{bz1} only
  in the presence of matter with $w \leq -3$.

\subsection {General analysis}

  All behaviors of $a(\tau) \to 0$ and $r(\tau)$ other than those enumerated
  in items 1-4 of the previous subsection either correspond to singular
  horizons (in the sense explained above) or to curvature singularities. A
  curvature singularity is avoided if [see (\ref{Riem-t}), the dot denotes
  $d/d\tau$]
\beq                                                         \label{KS-reg}
  |{\ddot a}/a|,\quad\ |{\dot a}{\dot r}/a|,\quad\ |\ddot r|
                    \ \ \ {\rm are\ finite.}
\eeq

  It is still of interest to analyze tidal forces acting on different
  geodesic observers in more general metrics with $a\to 0$, satisfying
  (\ref{KS-reg}), and to compare the results with the above horizon
  regularity conditions.

  Let us separately consider horizons with different asymptotics,
  generalizing the expressions (\ref{hor1})--(\ref{hor3}).

  First, assuming a horizon at finite $\tau=\tau_h$ and
  $a(\tau)\sim(\tau-\tau_h)^m$, the condition $|\ddot a/a| < \infty$
  immediately leads to $m=1$, hence $A(u) \sim u$, it is a first-order
  horizon ($n=1$). We can put
\beq
       a(\tau) \sim \tau-\tau_h, \cm                        \label{hor4}
       r(\tau) \approx r_h + C(\tau-\tau_h)^s,
\eeq
  with some constants $C$ and $s$ where $s\geq 2$ by virtue of (\ref{KS-reg}).
  Regular horizons, as we know from (\ref{hor1}), correspond to $s = 2k$,
  $k \in \N$.

  Second, for infinite $\tau$ at the horizon, we can take by analogy with
  (\ref{hor2})
\beq                                                        \label{hor5}
     \tau\to\pm\infty,\cm  a(\tau) \sim \e^{-m|\tau|}, \cm
       r(\tau) \approx r_h + C\e^{-s|\tau|}, \cm m > 0, \ \ s > 0.
\eeq
  with arbitrary constants $m$ and $s$. Regular horizons correspond to
  $s = km$, $k\in \N$.

  Third, in a similar way, we can take by analogy with (\ref{hor3})
\beq                                                        \label{hor6}
     \tau\to\pm\infty,\cm a(\tau) \sim |\tau|^{-m}, \cm
     r(\tau) \approx r_h + C|\tau|^{-s}, \cm m > 0, \ \ s > 0.
\eeq
  Regular horizons correspond to
\beq                                                        \label{m,s}
      m = \frac{n}{n-2}, \cm s = \frac{2k}{n-2}, \cm
                n = 3,4,\ldots, \ \ k \in \N.
\eeq

  Performing an analysis similar to that of Sec. IV, we arrive at Table 3.

\begin{table}  %%  III
  \caption{Types of horizons in KS cosmology}

\medskip
\centering
\renewcommand{\arraystretch}{1.15}
\tabcolsep 3pt
\begin{tabular}{|c|c|l|l|l|l|}
\hline
  No. &  $a,\ r$ & $m,\ s$
            & type by (\ref{crit}) & type by extensibility & $w$ \\
\hhline{|=|=|=|=|=|=|}
  1   & (\ref{hor4})  & $s=2$
        & usual or naked & regular, $n=1$  & $w=-3$  \\
\hline
  2   & (\ref{hor4})  & $2 < s < 4$
        & truly naked        & singular        & $-3 < w < -1$ \\
\hline
  3   & (\ref{hor4})  & $s\geq 4$
        & usual or naked     & regular, $n=1$, only if $2s=k\in\N$
                               & $w\leq -3$ \\
\hline
  4   & (\ref{hor5})  & $m>0,\ s=m$
        & usual or naked     & regular, $n=2$  & $w=-3$  \\
\hline
  5   & (\ref{hor5})  & $ m>0,\ m < s < 2m$
        & truly naked        & singular        & $-3 < w < -2$\\
\hline
  6   & (\ref{hor5})  & $ m>0,\ s \geq 2m$
        & usual or naked &
                        regular, $n=2$, only if $s/m  = k\in\N$
                               & $ w\leq -3$    \\
\hline
  7   & (\ref{hor6})  & $ m>1,\ s<2(m-1)$
        & truly naked        & singular  & $-3 < w <-\fract 53$\\
\hline
  8   & (\ref{hor6})  & $m>1,\ s= 2(m-1)$
        & naked              &
                    regular, $n\geq 3$, only if (\ref{m,s}) holds
                               & $ w=-3$   \\
\hline
  9   & (\ref{hor6})  & $m>1,\ s> 2 (m-1)$
        & usual              &
                    regular, $n\geq 3$, only if (\ref{m,s}) holds
                               & $w <-3$  \\
\hline
\end{tabular}

\end{table}

  As before, the presence of a TNH means that an observer moving along a
  timelike geodesic and reaching the horizon experiences infinite tidal
  forces. However, for a comoving observer, similar to a static observer
  in space-times considered above, the tidal forces [as well as all the
  curvature components (\ref{Riem-t})] are finite. The cosmological
  scenarios thus look differently for different groups of observers.
  Consider, for example. a TNH corresponding to line 2 of Table III.
  Then, an observer at rest reaches it at finite proper time and would be
  able to cross it if it were possible. However, an observer which follows a
  timelike geodesics with non-zero velocity with respect to the environment
  is subject to infinitely growing tidal forces and will be destroyed before
  reaching the horizon. This also applies to higher-order TNHs (lines 5 and
  7 of Table III). In the latter case, however, an observer at rest, for
  whom the tidal forces would be finite, never reaches the horizon since
  it would require an infinite time.

  It is also of interest to elucidate which kinds of matter are compatible
  with the space-times under discussion. Considering a non-interacting
  mixture of a vacuum fluid and an arbitrary kind of matter with the
  parameter $w = p_x/\rho \ne -1$ ($p_x = - T^x_x$ is the pressure in the
  $x$ direction; $w=-1$ corresponds to the vacuum fluid) in the manner of
  Sec.\,V.B and Ref.\,\cite{bz1}, we obtain the last column of Table III
  with appropriate values of $w$. All of them correspond to phantom matter.

  If there is no vacuum fluid, only simple regular horizons described by
  line 1 are possible, in agreement with the results of Sec.\,V.B and
  Ref.\,\cite{bz1}.

  A comparison shows that, for cosmological metrics, the parameter $w$
  behaves like $1/w$ for static metrics. This is not surprising since, in a
  sense, the energy density and the radial pressure interchange their
  roles when we change $A \to -A$. Curiously, according to Table III,
  cosmological metrics with TNHs require less exotic matter than those with
  a regular horizon.

  It is worth noting that the so-called nonscalar singularities (to which
  class the TNHs belong) in cosmology were also discussed in Sec.\,4.3 of
  \cite{e-sch77} and in \cite{ek,col}, where ``tilted cosmologies'' with
  perfect fluids were considered. The latter means that matter flows
  non-orthogonally to the homogeneity surfaces. A detailed comparison of
  our results with these papers is beyond the scope of this paper and
  deserves a separate treatment. Here, we only note that we do not restrict
  the matter to be necessarily a perfect fluid, and no ``tilting'' is
  assumed. In addition, we deal with KS \sph\ cosmologies whereas
  Ref.\,\cite{col} considered locally rotationally symmetric Bianchi-V models.

\section {Concluding remarks}

{\bf 1.} We have obtained the necessary conditions (\ref{gam'}), (\ref{r'})
  of a regular horizon, admitting a Kruskal-like extension, in terms of the
  general metric (\ref{ds}), and similar conditions (\ref{a-dot}),
  (\ref{r-dot}) for its cosmological counterpart, the general KS metric
  (\ref{KS-gen}). Comparing these general regularity conditions with
  the relations characterizing naked and truly naked horizons, we found
  that all TNHs and many horizons which seemed to be usual or simply naked
  represent singularities in the sense that geodesics terminate there at
  finite values of the canonical parameter.

  Our consideration was almost everywhere restricted to neighborhoods of
  (candidate) horizons, therefore we mostly spoke of naked, etc., horizons
  rather than \bhs. The results are equally applicable to any event, Cauchy
  or cosmological horizons in static or homogeneous \sph\ space-times in the
  framework of any metric theory of gravity. Moreover, the results obtained
  can be easily generalized to other space-time symmetries and dimensions.

\medskip\noi
{\bf 2.}
  In Sec.\,V.B, discussing matter with an arbitrary equation of state, we
  actually obtained that all kinds of horizons mentioned in Table I are
  possible but only in the presence of some ``vacuum fluid'' which may be
  represented by a cosmological constant or an electric or magnetic field.
  Meanwhile, to be compatible with any kind of horizon (including singular
  ones), the fluid itself must have negative radial pressure. This
  generalizes our previous result \cite{bz2} that only special kinds of
  fluids with particular values of the parameter $w = p_r/\rho$, including
  that with $w=-1/3$ (a fluid of disordered cosmic strings), are compatible
  with regular horizons.

  The examples considered show that naked and truly naked horizons do really
  appear in static solutions to the Einstein equations for certain kinds of
  matter, not to mention possible solutions of alternative theories of
  gravity. In particular, they may correspond to equations of state with
  $w < -1/3$ [see, e.g., (\ref{p/Lam})] often discussed in the context of
  non-phantom dark energy, or quintessence.

  In the cosmological context, the only kinds of matter compatible with
  horizons (both regular and singular) are vacuum fluids and phantom
  matter with $w = p_x/\rho< -1$, and a distinguished value compatible with
  regular horizons is $w = -3$. As in the static case, most of horizon types
  are only possible in the presence of a vacuum fluid.

\medskip\noi
{\bf 3.}
  In considering TNHs, we implicitly assumed that an extended body (or an
  observer) approaching a horizon and facing infinite tidal forces moves
  along a geodesic. Meanwhile, it is reasonable to ask what can happen in a
  more general situation. Is it possible to adjust the motion of an
  accelerated body in such a way that, even at a TNH, it could be subject to
  finite tidal forces unlike a geodesic observer?

  A direct inspection shows that it is really possible to avoid infinite
  tidal forces at a TNH by properly choosing a non-geodesic path. However,
  an inevitable price for this is that the acceleration experienced by such
  an observer tends to infinity when approaching the horizon. The
  calculations are carried out separately for the cases when the proper
  distance to the horizon is finite or infinite (the details will be
  presented elsewhere). The resulting infinite acceleration can be viewed as
  one more manifestation of a really singular nature of TNHs.

  An important point is, however, that a singular nature of many horizons in
  the sense of extensibility is not directly related to their ``truly
  naked'' character: we have seen that many horizons which do not create
  large or infinite tidal forces, are still inextensible. A reason for that
  evidently consists in the properties of matter in the near-horizon region,
  incompatible with analyticity of the functions $A(u)$ and $r(u)$.

\medskip\noi
{\bf 4.} A phenomenon of interest in KS cosmologies is the opportunity to
  receive information in the form of massive or massless particles from an
  epoch beyond a past horizon (if any), even though such a horizon could
  occur infinitely long ago according to our clocks. In reality, all such
  particles would be probably absorbed at a subsequent hot stage of the
  Universe evolution; but the existence of their traces in observable
  phenomena, e.g., in CMB properties, should not be excluded.

\medskip\noi
{\bf 5.}
  In the context of the early Universe, in addition to particle creation,
  it would be of interest to take into account one more quantum phenomenon,
  the dynamical Casimir effect related to the nontrivial topology of KS
  models, e.g., in the manner of Refs.\,\cite{NO99}, and its possible
  influence on the structure of singularities like those discussed in this
  paper.

  It would also be of interest to relate the origin of singularities in KS
  cosmology subject to quantum effects with their effective 2D description,
  in the manner of \cite{nooo99}, where quantum-corrected KS cosmologies were
  investigated. The effective 2D description makes the presentation
  qualitatively easier and may reveal a fundamental structure behind
  singularities, related to quantum effects.

\medskip\noi
{\bf 6.} The singular horizons in KS cosmology discussed in this paper (at
  least simple ones) may be considered as examples of the so-called
  finite-time singularities. Four types of such singularities are known and
  classified for isotropic FRW models in Ref.\,\cite{NOTs05}. Among them,
  the Big Rip (or type I singularity) is the most well-known and is widely
  discussed in connection with different models of dark energy. It is clear
  that in KS models, where we have two scale factors, such singularities may
  occur and their properties should be more diverse, and the corresponding
  classification should be naturally extended as compared with the
  one-scale-factor FRW cosmology. Such an extended classification, which
  should also apply to static, spherically symmetric analogues of KS
  cosmologies as well as to other, more complicated anisotropic cosmologies,
  is of significant interest. Let us mention some tentative results on this
  way.

  The Big Rip singularity in FRW models consists in an infinite
  growth of the scale factor $a(\tau)$ at finite proper time $\tau$,
  accompanied by $\rho \to \infty$, where $\rho$ is the density of matter
  with $w < -1$. Considering such a fluid, with $w = p_x/\rho = p_\bot/\rho
  < -1$ in the KS geometry (\ref{KS-tau}), we obtain that a Big Rip can also
  occur at finite $\tau = \tau_s$ but can be both isotropic and anisotropic:
  the volume factor $v$ and the density $\rho$ behave in the same way as in
  FRW models:
\beq                                                        \label{Rip-KS}
        v(\tau) = a(\tau)r^2(\tau) \sim  |\tau-\tau_s|^{2/(1+w)}, \cm
    \rho(\tau) \sim |\tau - \tau_s|^{-2},
\eeq
  but there can be an arbitrary ratio $N$ between the growth rates of
  $a(\tau)$ and $r(\tau)$, depending on initial conditions:
  $(\ln a)\dot{}/(\ln r)\dot{} = N =\const > 0$. In a similar way, we can
  discuss singularities of other types, as is done in Ref.\,\cite{NOTs05}.

  Static analogues of the Big Rip exist with $w = p_r/\rho = p_\bot/\rho$
  between zero and $-1$, in agreement with the above-mentioned
  $w \longleftrightarrow 1/w$ correspondence. An opportunity of interest is
  that $r\to r_s >0$ while $A \to \infty$ (a repulsive singularity on a
  sphere of finite radius) as the coordinate $u$ in (\ref{ds-u}) tends to a
  finite value $u_s$. The asymptotic behavior is
\beq                                                         \label{Rip-st}
     A \sim (u-u_s)^{-n}, \cm r \approx r_s + r_k(u-u_s)^k, \cm
     \rho\sim A^m,
\eeq
  for a fluid with $w = -1/(2m +1)$. Here, the exponents $k>1$, $n>0$ and
  $m >0$ (but $m \ne 1$) are connected by the relation $n(s-1) = 2-k$.

  Further details and other results on this subject will be presented
  elsewhere.

\subsection*{Acknowledgments}

  K.B. acknowledges partial financial support from
  Russian Basic Research Foundation Project 07-02-13614-ofi\_ts.
  E.E. and S.D.O. acknowledge support from MEC (Spain), projects
  FIS2006-02842 and 2005SGR-00790, and S.D.O. from MEC (Spain), project
  PIE2007-50/023.


\begin{thebibliography}{19}

\bibitem{HR1}
    G.T. Horowitz and S.F. Ross, \PRD {56} 2180 (1997).

\bibitem{HR2}
    G.T. Horowitz and S.F. Ross, \PRD {57} 1098 (1998).

\bibitem{ZP}
    V. Pravda and O.B. Zaslavskii, \CQG {22} 5053 (2005).

\bibitem{Z}
        O.B. Zaslavskii, \PRD {76} 024015 (2007).

\bibitem{wald}
    R. Wald, General Relativity (Univ. of Chicago Press, 1984).

\bibitem{tipler}
    F. J. Tipler, C. J. S. Clarke, and G. F. R. Ellis, in: ``General
    Relativity and Gravitation'', Vol. 2, pp. 97--206, Plenum, New
    York-London, 1980.

\bibitem{pois}
    E. Poisson and W. Israel, \PRL {63} 1663 (1989).

\bibitem{brady}
    P.R. Brady, in: ``Black Holes and Gravitational Waves'',
	Progr. Theor. Phys. Suppl. No. 136, 29--44 (1999).

\bibitem{nolan}
    B. C. Nolan, \PRD {60} 024014 (1999).

\bibitem{ori}
    A. Ori, \PRD {61} 064016 (2000).

\bibitem{craw}
        P. Aguiar and P. Crawford, Phys. Rev. D 62, 123511 (2000).

\bibitem{bd07}
    K.A. Bronnikov and I.G. Dymnikova, \CQG {24} 5803 (2007).

\bibitem{bz1}
    K.A. Bronnikov and O.B. Zaslavskii,
    Matter sources for a Null Big Bang, arXiv: 0710.5618;
    to appear in Class. Quantum Grav. (2008).

\bibitem{vac1}
    K.A. Bronnikov, Phys. Rev. D 64, 064013 (2001);

\bibitem{cold}
    K.A. Bronnikov, G. Cl\'ement, C. P. Constantinidis, and J. C. Fabris,
    Grav. Cosmol. 4, 128 (1998); Phys. Lett. A 243, 121 (1998).

\bibitem{walk70}
        M. Walker, J. Math. Phys. 11, 2280 (1970).

\bibitem{br79}
        K.A. Bronnikov.  Inversed  black  holes and anisotropic collapse.
    Izv. Vuzov SSSR, Fizika, No. 6, 32 (1979);
    Russ. Phys. J. {\bf 22}, 6, 594 (1979).

\bibitem{strobl}
    T. Kl\"osch and T. Strobl, \CQG {13} 2395--2422 (1996);
    {\bf 14}, 1689--1723 (1997).

\bibitem{kat}
    M.O. Katanaev, Nucl. Phys. Proc. Suppl. 88, 233 (2000), gr-qc/9912039;
    Proc. Steklov Inst. Math. 228, 158--183 (2000), gr-qc/9907088.

\bibitem{liber}
    S. Liberati, T. Rothman and S. Sonego, Phys. Rev. D 62, 024005 (2000).

\bibitem{e-sch77}
    G.F.R. Ellis and B.G. Schmidt, \GRG {8} 915 (1977).

\bibitem{bsh77}
    K.A. Bronnikov and G.N. Shikin,
        Izv. Vuzov SSSR, Fizika, No. 9, 25 (1977);
	Russ. Phys. J. {\bf 20}, 1138 (1977). 

\bibitem{GM}
    G.W. Gibbons, Nucl. Phys. B207, 337 (1982);\\
    G.W. Gibbons and K. Maeda, ibid., B298, 741 (1988).

\bibitem{GHS92}
    D. Garfinkle, G.T. Horowitz, and A. Strominger,
        Phys. Rev. D 43, 3140 (1991); ibid., 45, 3888 (1992).

\bibitem{bz2}
    K.A. Bronnikov and O.B. Zaslavskii, \PRD {78} 021501 (R) (2008).
    Arxiv: 0801.0889.

\bibitem{dym}
    I.G. Dymnikova, \GRG {24} 235 (1992);
        Class. Quantum Grav. {\bf 19}, 225 (2002);
        Int. J. Mod. Phys. {\bf D 12}, 1015 (2003).

\bibitem{br-NED}
    K.A. Bronnikov, Phys. Rev. D 63, 044005 (2001).

\bibitem{eliz1}
    A. Burinskii, E. Elizalde, S.R. Hildebrandt and G. Magli,
% {\sl Regular Sources of the Kerr-Schild class for Rotating and Nonrotating
% Black Hole Solutions},
        \PRD {65} 064039 (2002).  % gr-qc/0109085

\bibitem{eliz2}
    E. Elizalde and S.R. Hildebrandt,
% {\sl The family of regular interiors for non-rotating black holes with
% $T^0_0 = T^1_1$},
        \PRD {65} 124024 (2002).  % gr-qc/0202102

\bibitem{mann}
    K.C.K. Chan, J.H. Horne and R.B. Mann,
        Nucl. Phys. B 447, 441 (1995).

\bibitem{vac2}
    K.A. Bronnikov and G.N. Shikin, \GC {8} 107 (2002).

\bibitem{bu12}
    K.A. Bronnikov and J.C. Fabris, Phys. Rev. Lett. 96, 251101 (2001);\\
    K.A. Bronnikov, V.N. Melnikov and H. Dehnen,
        Gen. Rel. Grav. 39, 973 (2007).

\bibitem{ext}
    O.B. Zaslavskii, Class. Quantum Grav. 23, 4083 (2006).

\bibitem{bdd03}
    K.A. Bronnikov, A. Dobosz and I.G. Dymnikova, \CQG {20} 3797 (2003).

\bibitem{ek}
    G.F.R. Ellis and A.R. King, Comm. Math. Phys. {\bf 38}, 119 (1974).

\bibitem{col}
    C.B. Collins, Comm. Math. Phys. {\bf 39}, 131 (1974).

\bibitem{NO99}
    S. Nojiri and S.D. Odintsov, ,
    Int. J. Mod. Phys. A 15, 989 (2000), hep-th 9905089;
        \PRD {59} 044026 (1999), hep-th/9804033.

\bibitem{nooo99}
        S. Nojiri, O. Obregon, S.D. Odintsov and K.E. Osetrin,
    \PRD {60} 024008 (1999); hep-th/9902035.

\bibitem{NOTs05}
    S. Nojiri, S.D. Odintsov and S. Tsujikawa,
        \PRD {71} 063004 (2005); hep-th 0501025.


\end{thebibliography}
\end{document}